\documentclass[12pt,a4paper]{article}
\usepackage{amsmath}
\usepackage{float}
\usepackage{caption2}
\usepackage{graphicx}

\newcommand{\ra}{\ensuremath {\rangle} }

\begin{document}

\title{\textbf{\large{Hardy's nonlocality for generalized \textit{\textbf{n}}-particle GHZ states\thanks{This paper has been originally published in: Phys. Lett. A 327 (2004) 433-437.}}}}
\author{Jos\'{e} L.\ Cereceda\thanks{Electronic mail: jl.cereceda@telefonica.net} \\
\textit{C/Alto del Le\'{o}n 8, 4A, 28038 Madrid, Spain}}

\date{June 24, 2004}

\maketitle

\begin{abstract}
In this Letter we extend Hardy's nonlocality proof for two spin-1/2 particles [Phys. Rev. Lett. 71 (1993) 1665] to the case of $n$ spin-1/2 particles configured in the generalized GHZ state. We show that, for all $n\geq 3$, any entangled GHZ state violates the Bell inequality associated with the Hardy experiment. This feature is important since it has been shown [Phys. Rev. Lett. 88 (2002) 210402] that, for all $n$ odd, there are entangled GHZ states that do not violate any standard $n$-particle correlation Bell inequality.

\vspace{.3cm}
\noindent
\textit{PACS:} 03.65.Ud; 03.65.Ta
\vspace{.1cm}

\noindent
\textit{Keywords:} GHZ state; Hardy's nonlocality proof; Bell inequality violation

\end{abstract}

\vspace{.5cm}

In \cite{Hardy92} Hardy gave a proof of nonlocality without inequalities for two particles that only requires a total of four dimensions in Hilbert space. Shortly after the publication of this work, Pagonis and Clifton \cite{PC92} extended Hardy's theory to the case of $n$ spin-1/2 particles. However, all these proofs are only for particular entangled states. Later on, Hardy showed that, actually, the proof in \cite{Hardy92} can be carried out for any entangled state of two spin-1/2 particles except maximally entangled states \cite{Hardy93}. The converse of this result has been proved by Jordan \cite{Jordan94}: for any choice of two different measurement possibilities for each particle of a system of two spin-1/2 particles, a state can be found which admits Hardy's nonlocality. A simpler proof of this latter statement has been provided by Mermin \cite{Mermin94} and Kar \cite{Kar97a}. Kar also extended the converse of Hardy's result to a system of $n$ spin-1/2 particles \cite{Kar97b}. Wu and Xie \cite{WX96}, on the other hand, demonstrated Hardy's nonlocality theorem for almost all entangled states of three spin-1/2 particles by using a particular type of relationship among the coefficients of the given quantum state (see Eq.~(24) of \cite{WX96}). Subsequently, Ghosh et al. \cite{GKS98} proved Hardy's nonlocality for all really entangled states of three spin-1/2 particles, and Wu et al. \cite{WZP00} developed a Hardy-type nonlocality proof for the special case of three spin-1/2 particles in a maximally entangled GHZ state \cite{GHZ}.

In the present Letter we extend Hardy's original proof of nonlocality for two spin-1/2 particles \cite{Hardy93} to the case of $n$ spin-1/2 particles in the generalized GHZ state (see Eq.~(1)). It is shown that, for all $n\geq 3$, Hardy's nonlocality argument goes through for any entangled GHZ state, including a maximally entangled one. This contrasts with the case of $n=2$ particles, in which no maximally entangled state can exhibit Hardy's nonlocality. At any event, as we shall see, it turns out that the maximal amount of violation of the Bell inequality associated with the Hardy experiment (see Eq.~(7)) decreases exponentially with $n$. Nonlocality, however, remains manifest as $n\to\infty$. It is only when the number of particles becomes strictly infinite that the nonlocality argument breaks down. It will be shown that, for the Hardy experiment and for the class of states considered, the maximal discrepancy between the notion of local realism and quantum mechanics is obtained for the maximally entangled state of three particles. A remarkable feature of the Bell inequality (7) is that it can be violated by \textit{any\/} entangled GHZ state (1) for all $n\geq 3$. This feature is important because, as was recently shown in Ref.~\cite{ZBLW02} (see also Ref.~\cite{SG01}), for all $n$ odd there are entangled states of the form (1) which satisfy all Bell inequalities for correlation functions involving two dichotomic observables per particle. We will turn to this issue at the end of the Letter.

Consider $n$ spin-1/2 particles ($n\geq 2$) in the generalized GHZ state given by
\begin{equation}
|\psi\ra = \alpha |v_1v_2\ldots v_n\ra + \beta |w_1w_2\ldots w_n\ra,
\end{equation}
where $\{|v_k\ra,|w_k\ra\}$ is an arbitrary orthonormal basis in the state space of the $k\/$th particle, $k=1,2,\ldots,n$. Without loss of generality, it will be assumed that $\alpha$ and $\beta$ are taken to be real and positive, with $\alpha^2+\beta^2=1$. Now consider the physical observables $U_k$ and $D_k$ with corresponding operators $\hat{U}_k=|u^{+}_{k}\ra\langle u^{+}_{k}|-|u^{-}_{k}\ra\langle u^{-}_{k}|$ and $\hat{D}_k=|d^{+}_{k}\ra\langle d^{+}_{k}|-|d^{-}_{k}\ra\langle d^{-}_{k}|$, respectively. The eigenvectors $|u^{\pm}_{k}\ra$ and $|d^{\pm}_{k}\ra$ are related to the original basis vectors $|v_k\ra$ and $|w_k\ra$ by
\begin{align}
|u^{+}_k\ra &= \cos\alpha_k |v_k\ra + e^{i\delta_k}\sin\alpha_k |w_k\ra, \\
|u^{-}_k\ra &= -e^{-i\delta_k}\sin\alpha_k |v_k\ra + \cos\alpha_k |w_k\ra, \\
|d^{+}_k\ra &= \cos\beta_k |v_k\ra + e^{i\gamma_k}\sin\beta_k |w_k\ra, \\
|d^{-}_k\ra &= -e^{-i\gamma_k}\sin\beta_k |v_k\ra + \cos\beta_k |w_k\ra.
\end{align}
For a given quantum state (1), the observables $U_k$ and $D_k$ are required to satisfy the following conditions (so-called \textit{Hardy's nonlocality conditions\/}) \cite{Kar97b,WX96,GKS98}:
\begin{gather}
P(D_1 U_2 U_3\ldots U_n |+++\ldots +) = 0,  \nonumber \\
P(U_1 D_2 U_3\ldots U_n |+++\ldots +) = 0,  \nonumber \\
P(U_1 U_2 D_3\ldots U_n |+++\ldots +) = 0,  \nonumber \\
\vdots\quad\quad\vdots\quad\quad\vdots  \\
P(U_1 U_2 U_3\ldots D_n |+++\ldots +) = 0,  \nonumber \\
P(D_1 D_2 D_3\ldots D_n |---\ldots -) = 0,  \nonumber \\
P(U_1 U_2 U_3\ldots U_n |+++\ldots +) > 0,  \nonumber
\end{gather}
where, for example, $P(D_1 U_2 U_3\ldots U_n |+++\ldots +)$ denotes the probability that a joint measurement of $D_1,U_2,U_3,\ldots,U_n$, on particles $1,2,3,\ldots,n$, respectively, gives the outcome $+1$ for each of them. The nonlocality argument based on the above $(n+2)$ equations is as follows \cite{Kar97b,WX96,GKS98}. From the first $n$ equations of (6), we can deduce the following $n$ statements: (1) If $D_1,U_2,U_3,\ldots,U_n$ are measured, then necessarily $D_1=-1$ if $U_2=U_3=\cdots =U_n=+1$; (2) If $U_1,D_2,U_3,\ldots,U_n$ are measured, then necessarily $D_2=-1$ if $U_1=U_3=\cdots =U_n=+1; \ldots; (n)$ If $U_1,U_2,U_3,\ldots,D_n$ are measured, then necessarily $D_n=-1$ if $U_1=U_2=\cdots =U_{n-1}=+1$. In addition to this, from the last equation of (6), we get the $(n+1)$th statement: there is a nonzero probability of obtaining the results $U_1=U_2=\cdots =U_n=+1$ in a joint measurement of $U_1,U_2,\ldots,U_n$. Now, assuming that the particles are space-like separated and by combining the above $(n+1)$ statements with the assumption of local realism, we are led to conclude that there must be a nonzero probability to obtain the results $D_1=D_2=\cdots =D_n=-1$. But this contradicts the $(n+1)$th equation of (6).

Not surprisingly, the following Bell-type inequality can be derived which involves the $(n+2)$ probabilities appearing in (6):
\begin{align}
P(U_1 U_2 & U_3 \ldots U_n |+++\ldots +) \leq  P(D_1 D_2 D_3\ldots D_n |---\ldots -) \nonumber \\
& +P(D_1 U_2 U_3\ldots U_n |+++\ldots +) + P(U_1 D_2 U_3\ldots U_n |+++\ldots +) \nonumber \\
& \qquad +\cdots + P(U_1 U_2 U_3\ldots D_n |+++\ldots +).
\end{align}
Thus if the GHZ state (1) satisfies all the conditions in (6), it will automatically violate the Bell inequality (7).

Let us search for the constraints imposed by the fulfillment of the Hardy's nonlocality conditions. In the first place it will be noted that, in order for the first $n$ probabilities in (6) to vanish, it is necessary that
\begin{align}
\gamma_1+\delta_2+\delta_3+\cdots+\delta_n &= m_1\pi,  \nonumber  \\
\delta_1+\gamma_2+\delta_3+\cdots+\delta_n &= m_2\pi,  \nonumber  \\
\delta_1+\delta_2+\gamma_3+\cdots+\delta_n &= m_3\pi,  \nonumber  \\
\vdots\quad\quad\vdots\quad\quad\vdots &  \\
\delta_1+\delta_2+\delta_3+\cdots+\gamma_n &= m_n\pi,  \nonumber
\end{align}
for some integers $m_i =0,\pm1,\pm2,\ldots$ $(i=1,2,\ldots,n)$. From (8), we immediately get
\begin{equation}
\gamma_1+\gamma_2+\cdots+\gamma_n =(m_1+m_2+\cdots+m_n)\pi-(n-1)(\delta_1+\delta_2+\cdots+\delta_n).
\end{equation}
On the other hand, in order for the $(n+1)$th probability in (6) to vanish, the sum $\gamma_1+\gamma_2+\cdots+\gamma_n$ must equally be an integer multiple of $\pi$. So, in view of (9), this can be accomplished by taking
\begin{equation}
\delta_1+\delta_2+\cdots+\delta_n = \frac{m_{n+1}}{n-1}\,\pi,
\end{equation}
for some integer $m_{n+1} =0,\pm1,\pm2,\ldots .$ Thus, Eq.~(9) can be written as
\begin{equation}
\gamma_1+\gamma_2+\cdots+\gamma_n = (m_1+\cdots+m_n-m_{n+1})\pi.
\end{equation}
For the choice of phases in (8), it is readily shown that the vanishing of the first $n$ probabilities in (6) is equivalent to the fulfillment of the following $n$ conditions:
\begin{gather}
\tan\beta_1 \tan\alpha_2 \tan\alpha_3\cdots\ \tan\alpha_n = (-1)^{m_1+1}(\alpha/\beta), \nonumber \\
\tan\alpha_1 \tan\beta_2 \tan\alpha_3\cdots\ \tan\alpha_n = (-1)^{m_2+1}(\alpha/\beta), \nonumber \\
\tan\alpha_1 \tan\alpha_2 \tan\beta_3\cdots\ \tan\alpha_n = (-1)^{m_3+1}(\alpha/\beta), \nonumber \\
\vdots\quad\quad\vdots\quad\quad\vdots\qquad  \\
\tan\alpha_1 \tan\alpha_2 \tan\alpha_3\cdots\ \tan\beta_n = (-1)^{m_n+1}(\alpha/\beta). \nonumber
\end{gather}
Analogously, for the choice of phases in (11), the vanishing of the $(n+1)$th probability in (6) is equivalent to having
\begin{equation}
\tan\beta_1 \tan\beta_2 \cdots\ \tan\beta_n = (-1)^{m_1+\cdots+m_n-m_{n+1}+n+1}(\beta/\alpha).
\end{equation}
Multiplying all $n$ equations in (12), we obtain
\begin{equation}
(\tan\beta_1\tan\beta_2\cdots\tan\beta_n)\,(\tan\alpha_1\tan\alpha_2\cdots\tan\alpha_n)^{n-1} = 
(-1)^{m_1+\cdots+m_n+n} (\alpha/\beta)^{n} ,
\end{equation}
and using (13) in (14), we obtain
\begin{equation}
(\tan\alpha_1\tan\alpha_2\cdots\tan\alpha_n)^{n-1} = (-1)^{m_{n+1}-1} (\alpha/\beta)^{n+1} .
\end{equation}
Now, assuming without loss of generality that $0\leq\alpha_k\leq\pi/2$ for each $k$, it turns out that the constraint in (15) can be satisfied by all $n$ by taking $m_{n+1}$ to be
\begin{equation}
m_{n+1} = +1 , \quad \text{for all $n$},
\end{equation}
so that,
\begin{equation}
\delta_1+\delta_2+\cdots+\delta_n = \frac{\pi}{n-1}.
\end{equation}
The choice in (16) is essentially unique in the following sense. In the first place, as is clear from (15), $m_{n+1}$ must be an odd integer if the constraint (15) has to be fulfilled for $n$ odd. This is so irrespective of the values taken by the variables $\alpha_k$. Furthermore, for $0\leq\alpha_k\leq\pi/2$, $m_{n+1}$ must equally be an odd integer in order that the constraint (15) can be fulfilled for $n$ even. The choice $m_{n+1}=+1$ then arises from the fact that, for a given quantum state (1) and for given values of the variables $\alpha_k$, it gives a maximized probability $P(U_1 U_2\ldots U_n |++\ldots +)$, as can easily be verified from Eqs.~(10) and (19). Thus, putting $m_{n+1}=+1$, the constraint (15) becomes
\begin{equation}
\tan\alpha_1\tan\alpha_2\cdots\tan\alpha_n = (\alpha/\beta)^{\frac{n+1}{n-1}},
\end{equation}
for both $n$ odd and $n$ even, and for $0\leq\alpha_k\leq\pi/2$.

\begin{figure}[t]
\vspace{-.7cm}
\begin{minipage}[t]{0.5\linewidth}
\centering
\includegraphics[width=3.1in]{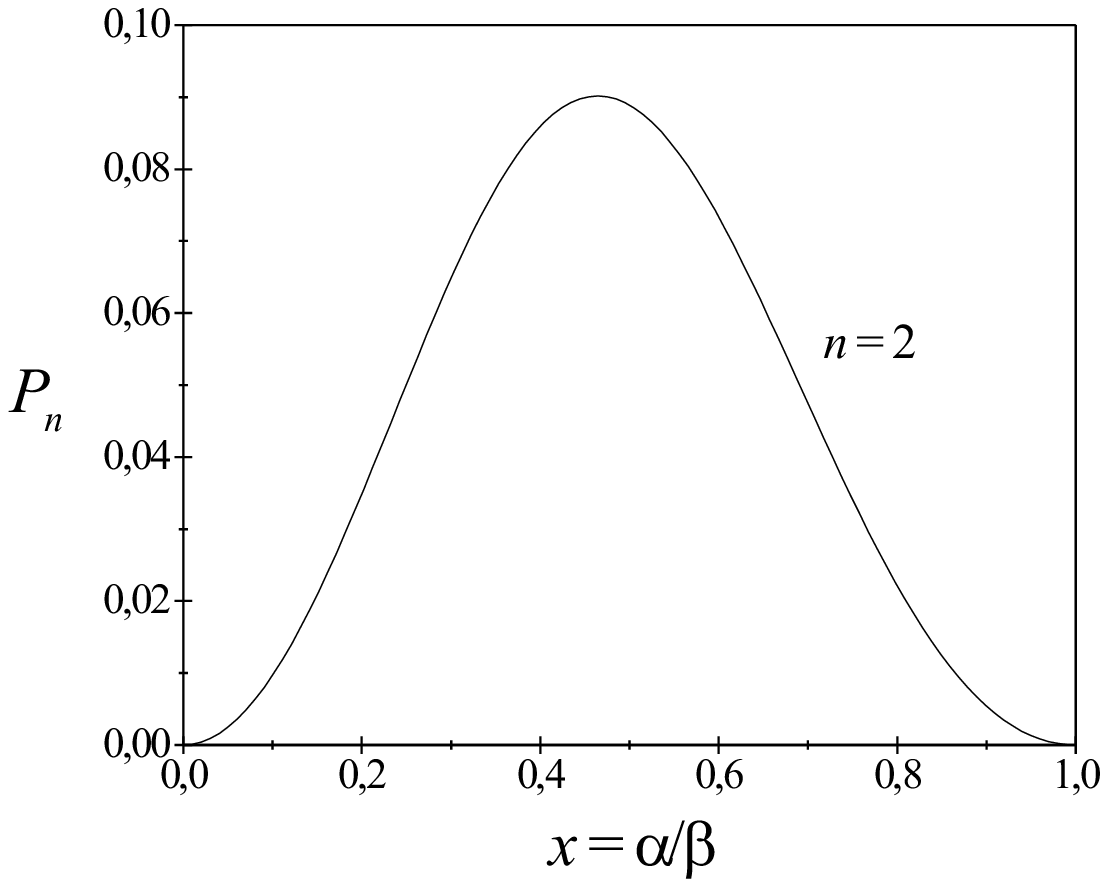}
\renewcommand{\figurename}{\footnotesize{Fig.}}
\renewcommand{\captionlabeldelim}{.~}
\setlength{\abovecaptionskip}{-.8cm}
\setcaptionmargin{.4cm}
\caption{\footnotesize{Plot of $P_n$ for $n=2$. The maximum value of $P_2$ is obtained for $\alpha/\beta =0.464$, and it is equal to $P^{\text{max}}_{2}=0.09$.}}  \label{fig1}
\end{minipage}
\begin{minipage}[t]{0.5\linewidth}
\centering
\includegraphics[width=3.1in]{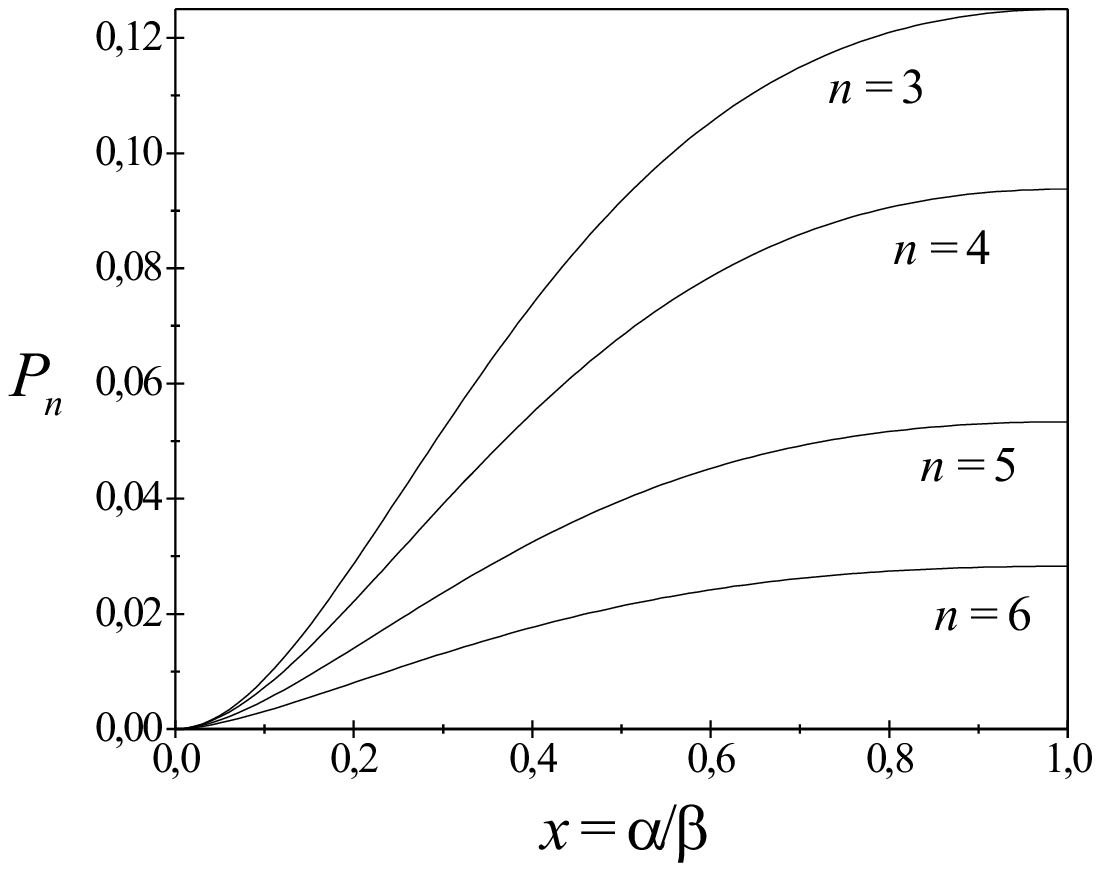}
\renewcommand{\figurename}{\footnotesize{Fig.}}
\renewcommand{\captionlabeldelim}{.~}
\setlength{\abovecaptionskip}{-.8cm}
\setcaptionmargin{.4cm}
\caption{\footnotesize{Plot of $P_n$ for $n=3,4,5,6$. For each $n\geq3$, $P_n$ reaches its maximum value for the maximally entangled state, \mbox{$\alpha/\beta =1$}.}}  \label{fig2}
\end{minipage}
\end{figure}

Let us look at the last probability in (6) (which we rewrite as $P_n$ for brevity). Explicitly, it is given by
\begin{equation}
P_n = \alpha^2 \prod_{k=1}^{n}\cos^2\alpha_k + \beta^2 \prod_{k=1}^{n}\sin^2\alpha_k
+ 2\alpha\beta \cos\left(\sum_{k=1}^{n}\delta_k\right) \prod_{k=1}^{n} \cos\alpha_k \sin\alpha_k .
\end{equation}
Clearly, the probability function (19) remains unchanged under the interchange of any pair of variables $\alpha_i$ and $\alpha_j$. This means that, in the configuration space spanned by the set of variables $\{\alpha_k\}$, the points giving an extremum value of $P_n$ must fulfill the condition that $\alpha_1=\alpha_2=\cdots=\alpha_n$. So, in what follows we assume that the variables $\alpha_k$ are taken such that $\alpha_1=\alpha_2=\cdots=\alpha_n\equiv\alpha_0$. The constraint in (18) then reads
\begin{equation}
\tan\alpha_0 = (\alpha/\beta)^{\frac{n+1}{n(n-1)}}.
\end{equation}
Thus, taking into account the relations in (17) and (20), we find
\begin{equation}
P_n = \frac{\alpha^2\beta^{2q} + \beta^2\alpha^{2q} 
 + 2\alpha^{q+1}\beta^{q+1}\cos(\frac{\pi}{n-1})}{(\alpha^p + \beta^p)^n},
\end{equation}
where
\begin{align}
p &= \frac{2(n+1)}{n(n-1)},   \\
q &= \frac{n+1}{n-1} .
\end{align}
For the first $n=2,3,4,5$, $P_n$ is given by
\begin{align}
P_2 &= \left(\frac{\alpha\beta^3 -\beta\alpha^3}{\alpha^3 +\beta^3}\right)^2 ,  \\
P_3 &= \frac{\alpha^2\beta^4 +\beta^2\alpha^4}{\left(\alpha^{\normalsize{\frac{4}{3}}}+\beta^{\frac{4}{3}}\right)^3}\, ,  \\
P_4 &= \frac{\alpha^2\beta^{\frac{10}{3}}+\beta^2\alpha^{\frac{10}{3}}+
\left(\alpha\beta\right)^{\frac{8}{3}}}{\left(\alpha^{\frac{5}{6}}+\beta^{\frac{5}{6}}\right)^4}\, ,  \\
P_5 &= \frac{\alpha^2\beta^3+\beta^2\alpha^3+
\sqrt{2}\left(\alpha\beta\right)^{\frac{5}{2}}}{\left(\alpha^{\frac{3}{5}}+\beta^{\frac{3}{5}}\right)^5}\, .
\end{align}
Alternatively, we can express $P_n$ as a function of $n$ and the ratio $x=\alpha/\beta$ as
\begin{equation}
P_n = \frac{x^2+x^{2q}+2x^{\frac{2n}{n-1}}\cos(\frac{\pi}{n-1})}
{(1+x^2)(1+x^p)^n}.
\end{equation}
The function in (28) is represented graphically in Fig.~\ref{fig1} for the case $n=2$, whereas it is plotted in 
Fig.~\ref{fig2} for the cases $n=3,4,5,6$. For concreteness, and without loss of generality, we have assumed that $\alpha\leq\beta$ in representing (28). (Please note that expression (28) remains invariant under the transformation $x\to 1/x$, so that the plot of $P_n(\beta/\alpha)$ for $\alpha\geq\beta$ looks exactly like the plot of $P_n(\alpha/\beta)$ for $\alpha\leq\beta$.)

A few additional remarks are in order regarding the probability function $P_n$. Firstly, for $n=2$, it reduces to that obtained by Hardy for two spin-1/2 particles (see Eq.~(20) of Ref.~\cite{Hardy93}), the value of $P_2$ being equal to zero whenever $\alpha=\beta$. Secondly, for $n\geq 3$, it is easily seen from (21) that $P_n>0$ for all values of $\alpha$ and $\beta$, except for $\alpha=0$ or $\beta=0$. Thus we have proved that any entangled GHZ state of three or more particles (including a maximally entangled one) can exhibit Hardy's nonlocality. Indeed, for each $n\geq 3$, it can be shown that $P_n$ reaches its maximum value for the maximally entangled state (see Fig.~\ref{fig2}). So, putting $x=1$ in (28), we find
\begin{equation}
P_n^{\text{max}}(n\geq3)= \left(\frac{1}{2^{n}}\right) \left[1+\cos\left(\frac{\pi}{n-1}\right)\right].
\end{equation}
From (29), we can see that $P_n^{\text{max}}(n\geq3)$ decreases exponentially with $n$. It should be noticed, however, that $P_n^{\text{max}}(n\geq3)$ or, more generally, the probability $P_n$ remains finite for a finite number of particles, with classical behavior only emerging \textit{discontinuously\/} in the (unrealizable) limit of an infinite number of particles \cite{PC92,PRC91,Cereceda95}. On the other hand, the absolute maximum value of $P_n$ is equal to $12.5\%$, which is realized for the maximally entangled state of three particles (cf.~Eq.~(29)). We note, incidentally, that this value agrees with that obtained in Refs.~\cite{GKS98,WZP00} for the particular case of three spin-1/2 particles in a maximally entangled GHZ state.

We conclude by noting a surprising result recently discovered by \.{Z}ukowski et al. \cite{ZBLW02} (see also Ref.~\cite{SG01}) according to which for all $n$ odd and for $\alpha\beta\leq 1/\sqrt{2^{n+1}}$ the correlations between the results of the measurements on $n$ particles in the generalized GHZ state (1) satisfy \textit{all\/} possible Bell inequalities for $n$-particle correlation functions, which involve two alternative dichotomic observables for each particle. Seemingly, this result contradicts our finding that, for all $n\geq3$, the Bell inequality (7) can be violated by \textit{any\/} entangled state (1). The explanation for this apparent contradiction stems from the fact that, actually, probability is the fundamental concept in any Bell experiment, and not the correlation function, so that one can derive correlations from probabilities, but the converse is not always possible \cite{CWKO03}. It is therefore deduced that the Bell inequality for probabilities (7) cannot be obtained from any standard Bell inequality for correlation measurements, in which local observers can choose between two dichotomic observables. In this respect, it should be added that Brukner et al. \cite{BLPZ} have recently derived Bell inequalities for correlation functions involving \textit{more than two\/} alternative measurement settings per observer, which are violated by the generalized GHZ state (1) for the full range of $\alpha$ or $\beta$ (see also Ref.~\cite{WZ03}). In particular, they have given a family of Bell inequalities for $n=3$ particles, which involve three measurement settings for the first two observers and two settings for the third one, and which are violated in the range of $0\leq \alpha\beta \leq 1/4$ by the factor $\sqrt{1+4\alpha^2\beta^2}$ \cite{BLPZ}. As the Bell inequality for probabilities (7) involves only two settings per observer, we conclude that, in a sense, the latter inequality is more efficient in detecting entanglement than the newly obtained Bell correlation inequalities for many measurement settings \cite{BLPZ}, thus underlining the basic role of the concept of probability for deriving Bell-type inequalities (see, in this respect, Ref.~\cite{CWKO03} for another Bell inequality for probabilities and two measurement settings per observer, and which is violated by all pure entangled states of three-qubit system).


\vspace{.5cm}

\begin{thebibliography}{99}

\bibitem{Hardy92} L. Hardy, Phys. Rev. Lett. 68 (1992) 2981.
\vspace{-.1cm}

\bibitem{PC92} C. Pagonis, R. Clifton, Phys. Lett. A 168 (1992) 100.
\vspace{-.1cm}

\bibitem{Hardy93} L. Hardy, Phys. Rev. Lett. 71 (1993) 1665.
\vspace{-.1cm}

\bibitem{Jordan94} T.F. Jordan, Phys. Rev. A 50 (1994) 62.
\vspace{-.1cm}

\bibitem{Mermin94} N.D. Mermin, Am. J. Phys. 62 (1994) 880.
\vspace{-.1cm}

\bibitem{Kar97a} G. Kar, J. Phys. A 30 (1997) L217.
\vspace{-.1cm}

\bibitem{Kar97b} G. Kar, Phys. Rev. A (1997) 1023.
\vspace{-.1cm}

\bibitem{WX96} X.-H. Wu, R.-H. Xie, Phys. Lett. A 211 (1996) 129.
\vspace{-.1cm}

\bibitem{GKS98} S. Ghosh, G. Kar, D. Sarkar, Phys. Lett. A 243 (1998) 249.
\vspace{-.1cm}

\bibitem{WZP00} X.-H. Wu, H.-S. Zong, H.-R. Pang, Phys. Lett. A 276 (2000) 221.
\vspace{-.1cm}

\bibitem{GHZ} D.M. Greenberger, M.A. Horne, A. Zeilinger, in: M. Kafatos (Ed.), Bell's Theorem, Quantum Theory,  and Conceptions of the Universe, Kluwer Academic, Dordrecht, 1989, p.~69.
\vspace{-.1cm}

\bibitem{ZBLW02} M. \.{Z}ukowski, C. Brukner, W. Laskowski, M.~Wie\'{s}niak, Phys. Rev. Lett. 88 (2002) 210402.
\vspace{-.1cm}

\bibitem{SG01} V. Scarani, N. Gisin, J. Phys. A 34 (2001) 6043.
\vspace{-.1cm}

\bibitem{PRC91} C. Pagonis, M.L.G. Redhead, R. Clifton, Phys. Lett. A 155 (1991) 441.
\vspace{-.1cm}

\bibitem{Cereceda95} J.L. Cereceda, Found. Phys. 25 (1995) 925.
\vspace{-.1cm}

\bibitem{CWKO03} J.-L. Chen, C.-F. Wu, L.C. Kwek, C.H. Oh, quant-ph/0311180.
\vspace{-.1cm}

\bibitem{BLPZ} C. Brukner, W. Laskowski, T. Paterek, M. \.{Z}ukowski, quant-ph/0303187.
\vspace{-.1cm}


\bibitem{WZ03} X.-H. Wu, H.-S. Zong, Phys. Lett. A 307 (2003) 262;

X.-H. Wu, H.-S. Zong, Phys. Rev. A 68 (2003) 032102.



\end{thebibliography}
\end{document}